\documentclass[12pt,a4paper]{article}
\usepackage{latexsym}
\usepackage{epsf}
\usepackage{amssymb}

\makeatletter
\@addtoreset{equation}{section}
\makeatother

 \def\unit{\hbox to 3.3pt{\hskip1.3pt \vrule height 7pt width .4pt \hskip.7pt
\vrule height 7.85pt width .4pt \kern-2.4pt
\hrulefill \kern-3pt
\raise 4pt\hbox{\char'40}}}

\def\half{{\textstyle {1 \over 2}}}
\def\quart{{\textstyle {1 \over 4}}}
\def\noverm#1#2{{\textstyle {#1 \over #2}}}
\def\Dpartial{{\cal D}}
\def\tr{\,{\rm tr}\,}
\def\makeatletter{\catcode`\@=11}
\makeatletter
\def\mathbox#1{\hbox{$\m@th#1$}}%
\def\math@ccstyles#1#2#3#4#5#6#7{{\leavevmode
      \setbox0\mathbox{#6#7}%
      \setbox2\mathbox{#4#5}%
      \dimen@ #3%
      \baselineskip\z@\lineskiplimit#1\lineskip\z@
      \vbox{\ialign{##\crcr
             \hfil \kern #2\box2 \hfil\crcr
             \noalign{\kern\dimen@}%
             \hfil\box0\hfil\crcr}}}}
\def\mathaccstyles{\math@ccstyles\maxdimen}
\def\maththroughstyles{\math@ccstyles{-\maxdimen}}
\def\unitmatrixDT%
 {\maththroughstyles{.45\ht0}\z@\displaystyle {\mathchar"006C}\displaystyle 1}
\begin{document}
\pagestyle{empty}
\begin{flushright}
\small
UG/00-16\\
VUB/TENA/00/08\\
{\bf hep-th/0011018}\\
November 1, 2000
\normalsize
\end{flushright}

\begin{center}


\vspace{.7cm}

{\Large {\bf Non-abelian Born-Infeld

                and

            kappa-symmetry}}

\vspace{.7cm}

 E.~A.~Bergshoeff${}^1$, M.~de Roo${}^1$ and A.~Sevrin${}^2$

\vskip 0.4truecm

 ${}^1$Institute for Theoretical Physics\\
   Nijenborgh 4, 9747 AG Groningen\\
     The Netherlands\\
\vskip 0.4truecm
 ${}^2$Theoretische Natuurkunde\\
   Vrije Universiteit Brussel\\
   Pleinlaan 2\\
   B-1050 Brussels, Belgium\\

\vskip 1.0cm


{\bf Abstract}

\end{center}

\begin{quotation}

\small

We define an iterative procedure to obtain a
non-abelian generalization of the Born-Infeld
action. This construction is made possible by the
use of the severe restrictions imposed by kappa-symmetry.
In this paper
we will present all bosonic terms in the action up to terms
quartic in the Yang-Mills field strength and all fermion
bilinear terms up to terms cubic in the field strength.
Already at
this order the fermionic terms do not satisfy the
symmetric trace-prescription.

\end{quotation}

\newpage

\pagestyle{plain}

\setcounter{section}{0}
\section{Introduction\label{Intro}}

One of the most intriguing features of D-branes is
their close connection with gauge theories.
Indeed, the effective theory describing the worldvolume dynamics of a
D$p$-brane is a $p+1$-dimensional field theory with, in the static gauge,
as bosonic degrees of freedom the transversal coordinates of the brane,
appearing as $9-p$ scalar fields, and the massless states of the open strings
ending on the brane which appear as a $U(1)$ gauge field.
When these fields vary slowly, the effective
action governing their dynamics is known to all orders in $\alpha '$.
It is the ten-dimensional
Born-Infeld action \cite{AT}, dimensionally reduced to $p+1$ dimensions.

Once several D-branes are present, the situation changes. The mass of the
strings stretching between two branes is proportional to the shortest
distance between the  branes. Starting off with $n$ well separated
D-branes we end up with a $U(1)^n$ theory, however, once the $n$ branes
coincide additional massless states appear which complete the gauge
multiplet to a non-abelian $U(n)$-theory \cite{EW1}. Contrary to the abelian case,
the effective action is not known to all orders in $\alpha '$. The first
term, quadratic in the field strength, is nothing but a dimensionally
reduced $U(n)$ Yang-Mills theory. The next order, which is quartic in the
field strength, was obtained from the four-gluon scattering amplitude
in open superstring theory \cite{GW} and
from a three-loop $\beta$-function calculation \cite{BP}. Based upon these
results and  other considerations, an all order proposal was
formulated for the effective action \cite{AT2}: the non-abelian
Born-Infeld action assumes essentially the same form as the abelian one,
however, all Lie algebra valued objects have to be symmetrized first before
taking the trace.  Other trace prescriptions, involving
commutators, have been given as well
\cite{AN}.
More recently, it was found that the symmetric trace prescription could
not be correct as it did not reproduce the mass spectrum of certain
D-brane configurations \cite{HT}, \cite{DST}.
It was shown in \cite{B}
that by adding commutator terms to the action the problem
might be cured. Indeed, as was pointed out in \cite{AT}, the notion of an
effective action for slowly varying fields is subtle in the non-abelian
case. In the effective action  higher derivative terms are dropped.
However because of
\begin{eqnarray}
D_i D_j F_{kl}=\frac 1 2 \{D_i , D_j\}F_{kl}-\frac i 2 {[}F_{ij}, F_{kl}
{]},
\end{eqnarray}
this is ambiguous. The analysis of the mass spectrum seems to indicate
that the symmetrized product of
derivatives acting on a field strength should be viewed as an acceleration
term which can safely be neglected, while the anti-symmetrized products
should be kept.
A systematic study of the $F^6$ terms \cite{STT} showed that using the
mass spectrum as a guideline, almost all terms at this order could be
determined. However due to the specific choices of backgrounds made in
\cite{STT}, certain terms do not contribute to the mass spectrum and as
a result can not be fixed in this way. A direct calculation from a
six-point open superstring amplitude or a five-loop $\beta$-function seems
unfeasible, so another approach is called for.

Till now, we ignored the fermionic degrees of freedom in our discussion.
The fully covariant worldvolume theory of a single D-brane in a type
II theory can
be formulated in terms of the following worldvolume fields\footnote{In
this paper $\mu,\nu=0,\ldots,9$ are spacetime indices, indices
$i,j=0,\ldots,p$ label the worldvolume coordinates $\sigma^i$.}: the
embedding coordinates $X^\mu(\sigma)$ (of which only the transverse
coordinates represent physical degrees of freedom), the Born-Infeld vector
field $V_i(\sigma)$, and $N=2$ spacetime fermionic fields
$\theta(\sigma)$.
In a curved background the D-brane can be coupled to the corresponding
type II supergravity superfields, and $N=2$ supersymmetry is realised
locally. In a flat background there is $N=2$ global supersymmetry.
This worldvolume theory has a local $\kappa$-symmetry,
which acts on the fermions as
\begin{equation}
\label{dt}
 \delta\bar\theta(\sigma) = \bar\kappa(\sigma)\,(1+\Gamma)\,,
\end{equation}
where $\Gamma$, which depends on worldvolume as well as background fields,
satisfies
\begin{equation}
  \Gamma^2 = \unitmatrixDT\,.
\end{equation}
The projection (\ref{dt})
makes it possible to gauge away half the fermionic
degrees of freedom. The field content then corresponds in a static gauge
to that of a supersymmetric Yang-Mills theory in $p+1$ dimensions.
There is still $N=2$ supersymmetry, but half of this is realised
nonlinearly.  These covariant D-brane actions have been constructed
in flat  \cite{Aga}, as well as in curved backgrounds
\cite{Ced,BT}. This paper examines the
suggestion in \cite{DST} that $\kappa$-symmetry might teach us
something about the ordenings appearing in the non-abelian Born-Infeld
action.

D-brane actions consist of the sum of a Born-Infeld term, coupling the
worldvolume fields to the NS-NS sector of the background,
and a Wess-Zumino term in which
the couplings to the R-R fields occur. Each part is separately
supersymmetric, but the two are related, in the abelian case,
by the $\kappa$-transformations. The
Wess-Zumino term is of a topological nature, and can therefore
be formulated in a metric-independent way. Its structure is
severely restricted, also in the non-abelian case. The Born-Infeld term
is much more complicated, and consequently its generalization from
the abelian to the non-abelian case is much more difficult. It is
natural to assume that also in the non-abelian case the
Born-Infeld and Wess-Zumino term are related by a
$\kappa$-symmetry. Our
aim is to use the knowledge of the Wess-Zumino term and the
properties of $\kappa$-symmetry to obtain information about the
non-abelian Born-Infeld term.

To construct this non-abelian generalization we
will use an iteration in the number of Yang-Mills field strengths
$F(V)$. In this paper
we will obtain all terms in the action up to and including the order $F^2$.
As we discussed
above, in the purely
bosonic terms the conflict between string theoretic
results and the symmetric trace prescription arises only at order
$F^6$, so it is clear that in this paper we will not
contribute to the discussion of these bosonic terms.
However, we find that in the fermionic sector already  at
quadratic order some of the fermionic terms
do not correspond to a symmetric trace.

This paper is organized as follows. We will discuss our choice of variables
in Section \ref{Fields}.
In Section \ref{Abel} we will define the iterative procedure and
illustrate it for the abelian case. In Sections \ref{YM}
and \ref{Summ} we derive and present our results
for the non-abelian case,
and bring them to gauge fixed form in Section \ref{Fix}.
In Section \ref{Concl}
we give our conclusions, and
point out a number of extensions and applications of this work.

We will end the Introduction by recalling briefly some
related work on supersymmetric D-brane actions. In four
dimensions the supersymmetrisation of the abelian Born-Infeld
action in $N=1$ supersymmetry has been known for a long time
\cite{CF}. More recently, this work has been extended
to the non-abelian Born-Infeld theory, and to $N=2$
supersymmetry \cite{K2,RTZ}. In particular, in \cite{K2} it
is remarked that $N=2$ supersymmetry in four dimensions is
not sufficient to resolve the ordering ambiguities, several
ordering precriptions give rise to supersymmetric actions.
So it seems that it is indeed necessary (and hopefully
sufficient) to consider $N=4$ in $D=4$, or, as in this
paper, the ten-dimensional supersymmetric Born-Infeld action
action. Several aspects of the ten-dimensional problem have
been studied in \cite{TR, M}. In particular, these authors
investigate the dependence of the action on transverse scalars,
where these scalars are generated by T-duality starting from the
D9-brane action. However, this is in the context of the symmetric
trace prescription.

\section{Worldvolume fields and transformations\label{Fields}}

The aim of this work is to obtain the effective action for $n$
overlapping Dp-branes, with $U(n)$ covariance on the worldvolume.
Before embarking on the construction, one has to carefully choose
the starting
point of the calculation. For  general Dp-branes  the situation is
complicated by the presence of the transverse scalar degrees of freedom,
which are in the adjoint representation of the Yang-Mills group. Not only
does one have to take commutators of these scalars into account, but also
the background fields will depend on these scalars \cite{D}. We avoid
these complications by limiting ourselves to the case of $n$ overlapping
D9-branes, and by choosing a  flat background. Through T-duality
D-branes for other values of $p$ can be obtained,
the extension to curved backgrounds
will be discussed in Section \ref{Concl}.

For $n$ overlapping D9-branes the completely gauge fixed result
should be the supersymmetric version of the non-abelian Born-Infeld
theory. Since the vector fields $V_i^A(\sigma)$, $A=1,\ldots,n^2$,
are in the adjoint representation of $U(n)$, we have to
make the same choice for the fermion fields $\theta$.
Therefore we start out with fields $\theta^A(\sigma)$,
which form a doublet
$(N=2)$ of Majorana-Weyl spinors for each $A$,
satisfying $\Gamma_{11}\theta^A=\theta^A$. After $\kappa$-gauge fixing
only half of each doublet will remain, and we have the
correct number of degrees of freedom for the supersymmetric Yang-Mills
theory.

This requires, that there are as many $\kappa$-symmetries as $\theta$'s,
so that the parameter of the $\kappa$ transformations will
have to be in the adjoint of $U(n)$ as well. Thus the $\theta^A$
transform as follows under
ordinary supersymmetry ($\epsilon$), $\kappa$-symmetry ($\kappa$),
Yang-Mills transformations $(\Lambda^A)$, and worldvolume
reparametrisations ($\xi^i$):
\begin{equation}
\label{dtall}
   \delta\bar\theta^A(\sigma) = -\bar\epsilon^A
     + \bar\kappa^B(\sigma)(\unitmatrixDT\,\delta^{BA} +
        \Gamma^{BA}(\sigma))
     + f^A{}_{BC}\Lambda^B(\sigma)\bar\theta^C(\sigma)
     + \xi^i(\sigma)\partial_i \bar\theta^A(\sigma)\,.
\end{equation}
Here $\epsilon^A$ are constant, $\Gamma^{AB}$ depends on the
worldvolume fields, and therefore on $\sigma$. It must satisfy
\begin{equation}
  \Gamma^{AB}\Gamma^{BC} = \delta^{AC}\unitmatrixDT\,.
\end{equation}
We will usually write these transformations in terms of
\begin{equation}
 \delta\bar\theta^A \equiv
 \bar\eta^A \equiv  \bar\kappa^B(\sigma)(\unitmatrixDT\,\delta^{BA} +
        \Gamma^{BA}(\sigma))\,.
\end{equation}

Useful information is obtained by considering commutators of
these transformations.
Because $\epsilon^A$ is constant we find from the commutator of
Yang-Mills and supersymmetry transformations that
\begin{equation}
      f^{A}{}_{BC}\Lambda^B\epsilon^C=0\to f_{ABC}\epsilon^C=0\,.
\end{equation}
Therefore $\epsilon = \epsilon^A T_A$, where $T_A$ are the
$U(n)$ generators, must be proportional to the unit
matrix, i.e., we can choose a basis in which there is only
one nonvanishing $\epsilon$ parameter.
So only a subset of the $\theta^A$ transform under supersymmetry,
and there is only one independent supersymmetry parameter.
The $\theta$'s which are presently
inert under supersymmetry will obtain their supersymmetry
transformations through $\kappa$-gauge fixing, as we shall see
in Section \ref{Fix}. {From} the commutator of $\kappa$-symmetry and
supersymmetry we find $\delta_\epsilon \eta^A = 0$. This implies that
\begin{equation}
  \delta_\epsilon \Gamma^{AB} = 0\,.
\end{equation}

The only scalars we have are  the embedding coordinates
$X^\mu(\sigma)$ for worldvolume directions.
There are several options that one could consider for the $X^\mu$:
\begin{enumerate}
\item We could assume that we are in the static gauge, i.e.,
\begin{equation}
    X^\mu(\sigma) = \delta^\mu_i\,\sigma^i\,,
\end{equation}
 from the
 beginning, so that the $X^\mu$ are absent. In this case there are no
 worldvolume reparametrisations, i.e., $\xi^i=0$ in (\ref{dtall}).
\item We could decide that the $X^\mu$ are in the singlet representation
 of the Yang-Mills group. The idea is that the $n$ branes overlap,
 there is only one set of worldvolume coordinates, and the corresponding
 reparametrization group would be sufficient to gauge fix a singlet
 set of embedding coordinates.
\item We could choose the $X^\mu$ in the adjoint representation of
 Yang-Mills in analogy with transverse coordinates for $p<9$.
 Here one thinks of starting with $n$ separate branes where each has its
 own worldvolume and embedding coordinates. When the branes overlap
 the embedding coordinates "fill up" to form elements of the adjoint
 representation.
 Clearly
 this requires a different approach toward the worldvolume
 reparametrisation invariance, which must then correspond to a sufficiently
 large symmetry group to gauge fix all these embedding functions.
\end{enumerate}
We have investigated the first two possibilities in the non-abelian case,
and we have found that only the first approach is consistent with the
iterative procedure that we employ. In Section \ref{YM} we will point out
where the first two choices start to diverge, in Section \ref{Concl}
we will briefly come back to the third possibility.

The transformation rules of the bosonic fields $V_i^A$, and, in case of
the second choice above, of the $X^\mu$, are determined iteratively
by requiring invariance of the action.

A special case of a $U(n)$ invariant non-abelian D-brane action is
of course the truncation to $U(1)^n$. In this case we know the answer:
a $\kappa$-invariant action is given by the sum of $n$ abelian
$D$-brane actions. This special case will be discussed again, since it
plays a role in making a choice between the different
possibilities for the variables $X^\mu$ discussed above.

Throughout this paper we will limit ourselves to terms in the action
and transformation rules which are
at most quadratic in the fermion fields.

\section{The iterative procedure and the abelian example\label{Abel}}

In this Section we will set up our iterative procedure and illustrate
it for the abelian case. To do this, we must first give some details
of the effective D9-brane action in a flat background \cite{Aga}.
In this case we can use a covariant formulation with embedding
coordinates $X^\mu$, spacetime fermions $\theta$, and the
Born-Infeld vector $V_i$. They transform under
supersymmetry $(\epsilon)$, $\kappa$-symmetry, worldvolume
reparametrisations $(\xi^i)$ and Maxwell gauge transformations
$(\Lambda)$ as:
\begin{eqnarray}
\label{dXab}
   \delta X^\mu  &=& \half\,\bar\epsilon\,\Gamma^\mu\theta
       + \half\,\bar\eta\,\Gamma^\mu\theta + \xi^i\partial X^\mu\,,
    \nonumber\\
\label{dth}
    \delta\bar\theta &=& -\bar\epsilon + \bar\eta +
            \xi^i\partial_i\bar\theta\,,
     \nonumber\\
\label{dV}
    \delta V_i &=& -\half(\bar\epsilon +\bar\eta)\gamma_i\theta
      + \xi^j F_{ji} + \partial_i(\Lambda + \xi^j V_j) \,,
\end{eqnarray}
with
\begin{equation}
\label{etakappa}
  \bar\eta = \bar\kappa (1+\Gamma)\,,\qquad (\Gamma)^2=\unitmatrixDT\,,
\end{equation}
and
\begin{equation}
   \gamma_i \equiv \Gamma^\mu\partial_i X^\mu \,.
\end{equation}

The Born-Infeld contribution reads
\begin{eqnarray}
{\cal L}_{\rm BI} &=& - \sqrt{-\det \,(g + {\cal F})}  \nonumber\\
     \label{LBI}
       &=& - \sqrt{-\det g}\,( 1+\noverm{1}{4} {\cal F}_{ij}{\cal F}^{ij}
        +\cdots)\,,
\end{eqnarray}
where in the second line we expand to second order in ${\cal F}$, which is
given by
\begin{eqnarray}
  {\cal F}_{ij} = F_{ij}(V) - B_{ij} \,.
\end{eqnarray}
In a flat background and in the quadratic fermion approximation,
 $B_{ij}$ is
\begin{equation}
\label{Bab}
  B_{ij} = - \bar\theta\sigma_3\gamma_{[i}\partial_{j]}\theta\,.
\end{equation}
The worldvolume metric reads
\begin{equation}
  g_{ij} = \eta_{ij} + \bar\theta \gamma_{(i}\partial_{j)}\theta\,,\qquad
  \eta_{ij} \equiv \partial_i X^\mu \partial_j X_\mu\,.
\label{geta}
\end{equation}
The metric $g$ and ${\cal F}$ are invariant under supersymmetry and transform
covariantly under $\kappa$-transformations.
The most useful form for comparison with the
non-abelian case is the expansion of (\ref{LBI})
to second order in fermions:
\begin{eqnarray}
\label{LBIquad}
    {\cal L}_{{\rm BI}} &=& -\sqrt{-\det\eta}\,\bigg(
    1 + \half\bar\theta\gamma^i\partial_i\theta +
         \half \bar\theta\sigma_3\gamma_{[i}\partial_{j]}\theta\, F^{ij}
  \nonumber\\
  &&\qquad  + \noverm{1}{4} F^{ij} F_{ij}
                 + \noverm{1}{2}\, \bar\theta\gamma_i\partial_j\theta\,T^{ij}
                   +\cdots\bigg)\,,
\end{eqnarray}
where $T^{ij}$ is the energymomentum tensor of the vector field:
\begin{equation}
   T^{ij} = F^{ik}F_{k}{}^j +\quart \eta^{ij} F_{kl}F^{kl}\,.
\end{equation}

The Wess-Zumino term takes on the following form:
\begin{eqnarray}
\label{LWZ}
  {\cal L}_{{\rm WZ}} &=&
   \epsilon^{i_1\ldots i_{10}} \times
   \sum_{k=0}^4   {(-1)^k\over 2^{k+1}\,k! (9-2k)!}
   \times
\nonumber\\
&& \times
  \,\bar\theta   \,
   {\cal P}_{(k)} \gamma_{i_1\ldots i_{9-2k}}\partial_{i_{10-2k}}\theta\,
    ({\cal F}^k)_{i_{11-2k}\ldots i_{10}}\,,
\end{eqnarray}
where
\begin{equation}
  {\cal P}_{(k)} =  \sigma_1\ ({\rm for}\ k=0,2,4)\,,
  \quad{\cal P}_{(k)} = i\sigma_2\ ({\rm for}\ k=1,3)\,.
\end{equation}
Note that the sum in (\ref{LWZ}) runs only to $k=4$, since the
RR-scalar field vanishes in the flat background.

It will be useful, also for the non-abelian case, to discuss why we make
this particular choice for the ${\cal P}_{(k)}$. The ${\cal P}_{(k)}$ are
chosen such that the contributions to the Wess-Zumino term are not total
derivatives.
For odd $k$ this fixes ${\cal P}_{(k)}$ to be $i\sigma_2$.
For even $k$ we could also have chosen $\unitmatrixDT$ or $\sigma_3$.
When we start looking at the iterative procedure later in this Section,
we will find that we need
\begin{equation}
  \{{\cal P}_{(0)}, {\cal P}_{(1)} \} = 0\,,
\end{equation}
which excludes $\unitmatrixDT$ for $k=0$. We have in principle the possibility
to have either $\sigma_1$ or $\sigma_3$ (or both) for $k=0$, and taking
$\sigma_1$ is a choice of basis for the $N=2$ fermions.  Note that
$\sigma_3$ in (\ref{Bab}) is correlated with the choice for $\sigma_1$ in
the Wess-Zumino term: had we chosen  ${\cal P}_{(0)}=\sigma_3$ we
would have found $\sigma_1$ in (\ref{Bab}).

The structure of (\ref{LWZ}) guarantees that the WZ-action
transforms into a total derivative under
the global supersymmetry transformation $\delta\bar\theta=-\bar\epsilon$.
Since we do not go beyond bilinear fermions we can use $F$ instead of
${\cal F}$ in the Wess-Zumino term.  The numerical coefficients in (\ref{LWZ})
are determined by $\kappa$-symmetry once the normalization of $F$ and $\theta$
are fixed in the Born-Infeld term.

Let us now consider the iterative construction of $\kappa$-symmetry.
The variation of the D-brane action takes on the form
\begin{equation}
\label{varkappa}
  \delta{\cal L} = - \bar\eta\, (1-\Gamma) {\cal T}\,.
\end{equation}
It indeed vanishes if $\eta$ is given by (\ref{etakappa}).
This variations have the following source:
\begin{equation}
   \delta{\cal L}_{{\rm BI}} =  - \bar\eta\, {\cal T} \,,
   \quad
    \delta{\cal L}_{{\rm WZ}} =   \bar\eta \,\Gamma\, {\cal T}\,.
\end{equation}
The variation of the Wess-Zumino term, together with the information that
$\Gamma^2=\unitmatrixDT$, is sufficient to determine both $\Gamma$ and ${\cal T}$
iteratively. Since ${\cal T}$ determines the variation of the Born-Infeld
term  this information is sufficient to obtain iteratively the
Born-Infeld part of the action.

The iteration is obtained by expanding $\Gamma$ and ${\cal T}$ in $F$:
\begin{eqnarray}
  \delta{\cal L} &=& -\bar\eta\,(1-(\Gamma_0+\Gamma_1+\cdots))
       ({\cal T}_0 + {\cal T}_1  +\cdots ) \nonumber\\
       &=& - \bar\eta\, ({\cal T}_0 - \Gamma_0{\cal T}_0
         + {\cal T}_1 - \Gamma_1{\cal T}_0 - \Gamma_0{\cal T}_1  +\cdots)\,,
\end{eqnarray}
where the indices indicate the order in $F$ and the $\Gamma_i$ satisfy
various identities which follow from $\Gamma^2=\unitmatrixDT$.
Since it will be useful to have the abelian results at hand for comparison
with the non-abelian calculation in Section \ref{YM}, we will work out the
beginning of this iteration in some detail.

Let us start with the order $F^0$. The contribution from (\ref{LWZ}) is
\begin{equation}
\label{LWZ0a}
  {\cal L}_{{\rm WZ}\,0} = {1\over 2\cdot 9!}\,
   \epsilon^{i_1\ldots i_{10}}\,\bar\theta
   \,\sigma_1
   \gamma_{i_1\ldots i_9}\partial_{i_{10}}\theta \,.
\end{equation}
For the $\kappa$-variation we only have to vary $\theta$ to obtain:
\begin{eqnarray}
\label{dLWZ0ab}
  \delta {\cal L}_{{\rm WZ}\,0} &=&
      {1\over 9!}\,\epsilon^{i_1\ldots i_{10}}\,\bar\eta \,\sigma_1
   \gamma_{i_1\ldots i_9}\partial_{i_{10}}\theta
   = \,\sqrt{-\det\eta} \,\bar\eta\, \sigma_1
     \Gamma^{(0)} \gamma^i\partial_i\theta \,.
\end{eqnarray}
Here $\Gamma^{(0)}$ is given by \footnote{We use both metrics $g$ and $\eta$.
They are defined in (\ref{geta}).}
\begin{equation}
\label{gamma0}
  \Gamma^{(0)} = {1\over 10!\sqrt{-\det g}}
   \epsilon^{i_1\ldots i_{10}}\gamma_{i_1\ldots i_{10}}\,,
\end{equation}
which satisfies
\begin{equation}
  \left(\Gamma^{(0)} \right)^2 = \unitmatrixDT\,.
\end{equation}
In (\ref{dLWZ0ab}) have used the property
\begin{equation}
\label{gamma0prop}
 \Gamma^{(0)} \gamma^{l_1\ldots l_k} =
   {(-)^{k(k-1)/2}\over (10-k)!\sqrt{-\det g}}
   \epsilon^{i_1\ldots i_{10-k}l_1\ldots l_k}\gamma_{i_1\ldots i_{10-k}}
\end{equation}
for $k=1$. {From} (\ref{dLWZ0ab}) we read off that
\begin{equation}
   \Gamma_0 = \Gamma^{(0)} \sigma_1\,,\qquad
   {\cal T}_0 =  \gamma^i\partial_i\theta\,.
\end{equation}
Obviously $\Gamma^2=\unitmatrixDT$ to this order.

So the Born-Infeld term should vary into ${\cal T}$, which is
indeed achieved by setting
\begin{equation}
\label{LBIab0}
  {\cal L}_{{\rm BI\,0}} = -\sqrt{-\det\eta} \bigg(
    1 + \half\,\bar\theta\gamma^i\partial_i\theta \bigg)\, .
\end{equation}
This gives
\begin{eqnarray}
  \label{dLBIab0}
  \delta{\cal L}_{{\rm BI}\,0}
   &=& -\sqrt{-\det\eta} \,\bar\eta\gamma^i\partial_i\theta\, ,
\end{eqnarray}
where we have used the variation of $X^\mu$ as given in (\ref{dXab}).

A similar analysis can be done for the terms of higher order in $F$.
At the linear level
the variation of the Wess-Zumino term is
\begin{eqnarray}
   \label{dLWZab1}
  \delta {\cal L}_{{\rm WZ\,1}} &=&
 \half \,\sqrt{-\det\eta}\,\bar\eta\,(i\sigma_2)\,
   \Gamma^{(0)}
   \big(\gamma^{jk}F_{jk}\gamma^i\partial_i\theta
        -2\gamma_i\partial_j\theta F^{ij}\big)\,.
\end{eqnarray}
The variation of the complete action should be
\begin{eqnarray}
   \delta{\cal L}_1 &=& -\sqrt{-\det\eta}\,
   \bar\eta\, \{ {\cal T}_1 - (\Gamma_0{\cal T}_1
           + \Gamma_1{\cal T}_0)\} \,.
\end{eqnarray}
So we read off that
\begin{equation}
   \Gamma_1 =  \Gamma_{(0)}\,(i\sigma_2)\half\gamma^{kl}F_{kl}\,,
   \qquad
    {\cal T}_1 = \sigma_3 \gamma_i\partial_j\theta F^{ij} \,.
\end{equation}
Note that $\Gamma^2=\unitmatrixDT$ at this order in $F$ because
\begin{equation}
   \{\sigma_1 \,,\ \Gamma_1\} = 0\,.
\end{equation}
This is a general feature: the  condition that $\Gamma^2=\unitmatrixDT$
only contains useful information at even orders in the expansion in
$F$. At odd orders it is just a consequence of the
properties of ${\cal P}_{(k)}$.

The variation under supersymmetry and $\kappa$-symmetry
of the term linear in $F$ in the Born-Infeld action
(\ref{LBIquad}) reads
\begin{eqnarray}
 \delta{\cal L}_{{\rm BI}\,1} &=&
  -\sqrt{-\det\eta}\,\bar\eta\,
      \sigma_3\gamma_i\partial_j\theta F^{ij}
   \nonumber\\
 &&\qquad \label{dLBIab11}
    - \half\,(\bar\epsilon+\bar\eta)\gamma_i\theta\, \partial_j \{
      \sqrt{-\det\eta}\, F^{ij}\}
    \,.
\end{eqnarray}
The first term is the required contribution of ${\cal T}_1$. The second
term must be cancelled by the variation of $V$ in the $F^2$ term.
The $F^2$ term gives
\begin{equation}
  -  \delta V_i \,\partial_j\{ \sqrt{-\det\eta}\, F^{ij}\}
\end{equation}
which implies the following variation of $V_i$:
\begin{equation}
\label{dVab1}
  \delta V_i = - \half\,(\bar\epsilon+\bar\eta)\sigma_3\gamma_i\theta \,.
\end{equation}
Therefore the combination
\begin{equation}
  {\cal F}_{ij} = F_{ij} +
               \bar\theta\sigma_3\gamma_{[i}\partial_{j]}\theta
\end{equation}
is supersymmetric and transforms covariantly under $\kappa$-symmetry.

At the quadratic level we get
\begin{eqnarray}
\label{dLWZab2}
  &&\delta {\cal L}_{{\rm WZ\,2}} =
       {1\over 8}\,\sqrt{-\det\eta} \,\bar\eta\,\sigma_1\,
    \Gamma^{(0)}
   \big(\gamma_{ijkl}\gamma^m\partial_m\theta
        -4\gamma_{ijk}\partial_l\theta \big)
        F^{[ij}F^{kl]} \,.
\end{eqnarray}
The order 2 terms in the variation of the total action are
\begin{eqnarray}
\label{termsab2}
  \delta{\cal L}_2 &=& -\sqrt{-\det\eta}\,\bar\eta
    \bigg( -\Gamma^{(0)}( \Gamma_2 \gamma^i\partial_i\theta
      +\Gamma_1{\cal T}_1 + \sigma_1{\cal T}_2)
      + \unitmatrixDT\,{\cal T}_2\bigg)\,,
\end{eqnarray}
where $\Gamma_1$ and ${\cal T}_1$ where determined at the linear level.
On the other hand, from $\Gamma^2=\unitmatrixDT$ we have
\begin{equation}
   \sigma_1\Gamma_2+\Gamma_2\sigma_1 + \Gamma_1\Gamma_1 = 0\,,
\end{equation}
from which we obtain (using $\Gamma_2\sim \sigma_1$)
\begin{eqnarray}
   \Gamma_2 &=& -\half \sigma_1\Gamma_1\Gamma_1
   \nonumber\\
    &=&  \sigma_1 \{\noverm{1}{8} \gamma_{ijkl}F^{ij}F^{kl}
       -\quart F^{kl}F_{kl}\}\,.
\end{eqnarray}
Substituting all this in (\ref{termsab2}), we find that
\begin{equation}
\label{T2ab}
  {\cal T}_2 = \gamma_i\partial_j\theta (F^{ik}F_k{}^j + \quart\eta^{ij}
          F^{kl}F_{kl})\,.
\end{equation}
This indeed agrees with the variation of the Born-Infeld action.

There is a feature about the abelian case which
just starts being visible in
the  quadratic terms. It is obviously possible to write $\Gamma$ at
this order in the form
\begin{equation}
  \Gamma = (1 - \quart F^{kl}F_{kl})
     \Gamma^{(0)}((\sigma_1+\half(i\sigma_2)
      \gamma^{kl}F_{kl}
      +\noverm{1}{8}\sigma_1\gamma_{ijkl}F^{ij}F^{kl})
\end{equation}
up to terms of higher order in $F$. In fact, this factorisation
is a general feature
of the abelian action which is also valid for the complete answer:
\begin{equation}
\label{fullgamma}
  \Gamma = {\sqrt{-\det g}\over \sqrt{-\det(g+{\cal F})}}
     \Gamma^{(0)}
     \sum_{k=0}^5 {1\over 2^k k!} {\cal P}_{(k)} \gamma^{i_1\ldots i_{2k}}
       {\cal F}^{k}_{i_1\ldots i_2k}\,.
\end{equation}
The iterative procedure will obviously confirm this factorisation, as
is shown by continuing to higher orders in $F$. However, in this
construction it is not clear why the factorisation should occur.
This is  an issue in the non-abelian situation where the
complete answer is not known. Note that the factor
\begin{equation}
     {\sqrt{-\det g}\over \sqrt{-\det(g+{\cal F})}}
\end{equation}
in (\ref{fullgamma}) contains the inverse of the Born-Infeld action.
The idea that $\Gamma$ provides the explicit form of the
Born-Infeld action was part of our motivation to use $\kappa$-symmetry
as a means of constructing the non-abelian Born-Infeld action.

{From} (\ref{T2ab}) it is clear that ${\cal T}$, at least at the quadratic
level, shows a similar factorisation property as $\Gamma$. The full
answer for ${\cal T}$ is of the form:
\begin{eqnarray}
   {\cal T} &=& \sqrt{-\det(g+{\cal F})}\,
   \sum_{k=0}^{\infty} (\sigma_3)^k \gamma_i\partial_j\theta
      ({\cal F}^k)^{ij}\,,
\end{eqnarray}
where
\begin{equation}
\begin{array}{ll}
   ({\cal F}^k)^{ij} = {\cal F}^{il_1}
      {\cal F}_{l_1l_2}\cdots {\cal F}^{l_{k-1}j} & k>0 \,, \\
   ({\cal F}^k)^{ij}= g^{ij}                  & k=0\,.
\end{array}
\end{equation}
We will see a similar feature in the non-abelian case.

The abelian case discussed above can easily be generalized to $U(1)^n$.
Then the $\kappa$-symmetric action is just the sum of $n$ actions
of the type discussed in this Section. For $n$ overlapping branes
one would need only one set of embedding coordinates $X^\mu$ to
describe this truncation of the non-abelian situation. This would be
similar to treating $X^\mu$ as a singlet of $U(n)$ in the non-abelian
case. Note however that this sum of actions is very different from
a single Born-Infeld action with
worldvolume metric
\begin{equation}
\label{metricUn}
   g_{ij} = \eta_{ij} + \bar\theta^A\gamma_{(i}\partial_{j)}\theta^A\,,
\end{equation}
summed over the $n$ $U(n)$ branes. A metric (\ref{metricUn}) would be
like taking the trace {\it inside} the root of the Born-Infeld term,
while it
is known from open
string amplitude calculations
that there should be a single trace (with some ordening
prescription) which produces in the $U(1)^n$ case a sum of
separate Born-Infeld terms.

\section{The non-abelian Born-Infeld action\label{YM}}

For the non-abelian case we start with the following Wess-Zumino term:
\begin{eqnarray}
  {\cal L}_{{\rm WZ}} &=&
   \epsilon^{i_1\ldots i_{10}} \times
   \sum_{k=0}^4   {(-1)^k\over 2^{k+1}\,k! (9-2k)!}
       \times
\nonumber\\
&&\label{LWZnab} \times
  \,\bar\theta^A \, {\cal P}_{(k)}^{AB\,C_1\ldots C_k} \,
  \gamma_{i_1\ldots i_{9-2k}}\Dpartial_{i_{10-2k}}\theta^B\,
      F_{i_{11-2k}i_{12-2k}}^{C_1}\cdots F_{i_9i_{10}}^{C_k}\,.
\label{WZterm}
\end{eqnarray}
The tensors ${\cal P}$ are symmetric in the indices $C_i$
contracted with  $F$. ${\cal P}$ also contains the Pauli matrices
to specify the $N=2$ structure for the fermions. We have
the following possibilites:
\begin{equation}
\begin{array}{lll}
    k\ {{\rm even:}} & \unitmatrixDT_2,\ \sigma_1,\ \sigma_3
                     & {{\rm symmetry\ in}}\ AB\,, \\
                     & i\sigma_2
                     & {{\rm antisymmetry\ in}}\ AB\,, \\
    k\ {{\rm odd:}}  & \unitmatrixDT_2,\ \sigma_1,\ \sigma_3
                     & {{\rm antisymmetry\ in}}\ AB\,, \\
                     & i\sigma_2
                     & {{\rm symmetry\ in}}\ AB \,.
\end{array}
\end{equation}
This requirement follows from the fact that the bilinear fermions
in the action (\ref{WZterm}) should not be a total derivative. The Yang-Mills
structure of ${\cal P}$ arises from the trace of $k+2$ generators
in the fundamental representation\footnote{The reason
that we may use this representation follows
from open string theory, where the gauge theory factors arise
from Chan-Paton factors in the fundamental representation.}
of $U(n)$, and
will be built from the structure constant
$f_{ABC}$ (completely antisymmetric) and from the completely
symmetric tensors $d_{ABC}$. In Appendix \ref{AppUn} we gather
useful properties of these tensors.

The general form of (\ref{LWZnab}) follows from the requirement that
the Wess-Zumino term is of topological nature, i.e., independent
of the metric. The coefficients have been chosen equal to those in the
abelian case, which amounts to a particular normalization of the
${\cal P}_{(k)}$. Note that we do not assume a particular ordening in
the trace, i.e., there is no a priori symmetry imposed between the indices
$C_i$ on the one hand, and $A,B$ on the other hand.

So at order 0 we start with\footnote{The choice for $\sigma_1$ is again
a choice of the $N=2$ basis, which we take in agreement with the
 abelian case.}
\begin{equation}
   {\cal P}_{(0)}^{AB} = \sigma_1 \tr T^AT^B = \sigma_1 \delta^{AB}\,,
\end{equation}
and the variation of the lowest order contribution in (\ref{LWZnab})
is
\begin{eqnarray}
\label{dLWZnab0}
 \delta {\cal L}_{{\rm WZ}\,0} &=&
      \sqrt{-\det \eta} \,\bar\eta^A \sigma_1
     \Gamma^{(0)} \gamma^i\Dpartial_i\theta^A \,.
\end{eqnarray}
We write the variation of the complete action at this order as
\begin{equation}
\label{dL0n}
  \delta {\cal L}_0 =  -\sqrt{-\det \eta}\, \bar\eta^A
    (\delta^{AB}
     - \Gamma_0^{AB}){\cal T}_0^B\,.
\end{equation}
so that
\begin{equation}
  \Gamma_0^{AB} = \Gamma^{(0)}\delta^{AB}\sigma_1\,,\qquad
   {\cal T}^A_0 = \gamma^i\Dpartial_i\theta^A\,.
\end{equation}
Cleary $\Gamma^{AC}\Gamma^{CB} = \delta^{AB}$\,.
In the above we considered $X^\mu$ to be a singlet under $U(n)$
transformations. To go to the static gauge we would have to set
$\det\eta\to -1$.

The Born-Infeld term must reproduce the first term in (\ref{dL0n}).
The only choice is to have
\begin{equation}
  \label{LBInab0}
    {\cal L}_{{\rm BI}} = -\sqrt{-\det\eta}\,\bigg(
    1 + \half\,\bar\theta^A\gamma^i\Dpartial_i\theta^A \bigg)\,.
\end{equation}
We indeed find the correct variation if we set
\begin{equation}
   \label{dXnab}
   \delta X^\mu  = \half\,\bar\epsilon^A\,\Gamma^\mu\theta^A
       + \half\,\bar\eta^A\,\Gamma^\mu\theta^A \,.
\end{equation}
With this choice of $\delta X$ the metric (\ref{metricUn})
becomes supersymmetric and covariant under $\kappa$-transformations.
As we explained at the end of Section \ref{Abel} this is not
the natural metric for the non-abelian (or for the $U(1)^n$)
situation. So the different choices for $X^\mu$
start to diverge at this point. At the linear level there is still
no crucial difference between the choice of a singlet $X^\mu$
and the static gauge. At the quadratic level, however, the
singlet choice will fail.

The variation of the linear contribution to the Wess-Zumino term gives:
\begin{eqnarray}
\label{dLWZnab1}
  &&\delta {\cal L}_{{\rm WZ\,1}} =
       \half\, \sqrt{-\det\eta}\,\bar\eta^A {\cal P}_{(1)}^{ABC}\,
    \Gamma^{(0)}
   \big(\gamma_{ij}\gamma^k\Dpartial_k\theta^B
        -2\gamma_i\Dpartial_j\theta^B \big)
        F^{C\,ij} \,.
\end{eqnarray}
The first term must correspond to $\Gamma_1{\cal T}_0$, from
which we read off that
\begin{equation}
   \Gamma_1^{AB} = \Gamma^{(0)} {\cal P}_{(1)}^{ABC}
     \half\gamma_{ij}F^{ij\,C}\,.
\end{equation}
{From} $\Gamma^2=\unitmatrixDT$ at linear order we find
\begin{equation}
    \{\sigma_1\,,\ \Gamma_1^{AB} \} = 0       \,,
\end{equation}
so that ${\cal P}_{(1)}$ must have the following form:
\begin{equation}
\label{P1def}
   {\cal P}_{(1)}^{ABC} = (i\sigma_2) d^{ABC} + c_1\sigma_3 f^{ABC}\,.
\end{equation}
The coefficient of the $d$-term is chosen to agree with the abelian case.
The coefficient of the $f$-term is arbitrary,
and with a field redefinition
\begin{equation}
   \theta^A\to \theta^A -
       \quart c_1 (i\sigma_2)f^{ABC}\,\gamma^{kl}\theta^BF_{kl}^C,
\end{equation}
and a corresponding redefinition of the vector field,
the $f$-term can be eliminated. This is the choice we
will make, because then we stay as close as possible
to the abelian situation.

In fact, the whole term linear in $F$ in the Wess-Zumino action
can be transformed away by a field redefinition, also
in the abelian case. It is not surprising that
these linear terms can be eliminated, since they are
part of the supersymmetrisation of the bosonic $F^3$ term,
which we know to be absent. The reason we will keep
the usual linear term is that in this form the answer
in the abelian case
takes on a relatively simple form \cite{Aga}.

We also find
\begin{equation}
  {\cal T}_1^A = -\sigma_1{\cal P}_{(1)}^{ABC}
   \gamma_i\Dpartial_j\theta^B F^{ij\,C}\,,
\end{equation}
so that
in analogy with the abelian case the Born-Infeld term must
contain:
\begin{equation}
\label{LBI1nab}
   {\cal L}_{{\rm BI\,1}} =   - \sqrt{-\det\eta}\,\bigg(
   - \half\, \bar\theta^A \sigma_1 {\cal P}_{(1)}^{ABC}
       \gamma_{[i}\Dpartial_{j]}\theta^B\, F^{ij\,C}
      + \noverm{1}{4} F^{ij\,A} F_{ij}^A \bigg)\,.
\end{equation}
The variation of this term reproduces correctly the ${\cal T}_1$
contribution, and the remainder is cancelled by introducing
a variation of $V_i$:
\begin{equation}
   \delta V_i^C = + \half\,(\bar\epsilon^A+\bar\eta^A)\,
           \sigma_1{\cal P}_{(1)}^{ABC}\gamma_i\theta^B \,.
\end{equation}
We can then define a supersymmetric and $\kappa$-covariant
${\cal F}^C$ as
\begin{equation}
\label{calF}
  {\cal F}_{ij}^C = F_{ij}^C - \bar\theta^A \sigma_1 {\cal P}_{(1)}^{ABC}
                 \gamma_{[i}\Dpartial_{j]}\theta^B\,.
\end{equation}
This defines the non-abelian generalization of the NS-NS two-form field,
to this order.

At the quadratic level things become more complicated. The variation
of the Wess-Zumino term gives:
\begin{eqnarray}
  \delta {\cal L}_{{\rm WZ\,2}} &=&
        \sqrt{-\det\eta}\,\bar\eta^A {\cal P}_{(2)}^{ABCD}\,
         \Gamma^{(0)}\times \nonumber\\
    \label{dLWZnab2}
  && \times \{\,\noverm{1}{8}\gamma_{ijkl}\gamma^m\Dpartial_m\theta^B
        - \half\gamma_{ijk}\Dpartial_l\theta^B \}\,
        F^{C\,ij}F^{D\,kl}\,.
\end{eqnarray}
The variations quadratic in $F$ will have to generate the following
contributions:
\begin{eqnarray}
  \delta{\cal L}_2 &=& - \sqrt{-\det\eta}\,\bar\eta^A
    \bigg( -( \Gamma_2^{AB} \gamma^i\Dpartial_i\theta^B
      +\Gamma_1^{AB}{\cal T}_1^B + \Gamma_0^{AB}{\cal T}_2^B)
\nonumber\\
   &&\qquad
   \label{terms2}
      +\ \unitmatrixDT\,{\cal T}_2^A\bigg)
\end{eqnarray}
We can determine $\Gamma_2$ from the requirement that
$\Gamma^2=\unitmatrixDT$ at order 2:
\begin{equation}
\label{gamma2}
  \sigma_1\Gamma_2^{AB} + \Gamma_2^{AB}\sigma_1
         + \Gamma_1^{AC}\Gamma_1^{CB} = 0\,.
\end{equation}
This calculation assumes that $[\Gamma_2\,,\ \sigma_1]=0$, and
requires the product of the ${\cal P}_{(1)}$-tensors.
The result is
\begin{eqnarray}
   \Gamma_2^{AB} &=& - \Gamma^{(0)} \sigma_1\,\bigg(
    {\cal S}^{ABCD} \big( \noverm{1}{8} \gamma_{ijkl}
          F^{ij\,C} F^{kl\,D} -\quart F_{kl}^C F^{kl\,D} \big)
       \nonumber\\
    &&\quad  +  {\cal A}^{ABCD}
           \half \gamma_{ij} F^{ik\,C} F_k{}^{j\,D} \bigg) \,.
\end{eqnarray}
Here we have defined
\begin{eqnarray}
   {\cal S}^{ABCD} &\equiv& {\cal P}^{AE(C}_{(1)} {\cal P}^{BD)E}_{(1)} =
              - d^{AE(C}d^{BD)E}\,,
   \\
    {\cal A}^{ABCD} &\equiv& {\cal P}^{AE[C}_{(1)} {\cal P}^{BD]E}_{(1)} =
              - d^{AE[C}d^{BD]E}\,,
\end{eqnarray}
where the (anti-)symmetrization is over the indices $C$ and $D$
only. Note that  tensors ${\cal S}^{ABCD}$ and ${\cal A}^{ABCD}$ are
then symmetric and antisymmetric, resp., in the index pairs $AB$ and $CD$.

To solve the remainder of (\ref{terms2}) we have to make the choice
\begin{equation}
   {\cal P}_{(2)}^{ABCD} = -\sigma_1{\cal S}^{ABCD}\,.
\end{equation}
We then find the following result for ${\cal T}_2$:
\begin{eqnarray}
   {\cal T}_2^A &=& -{\cal S}^{ABCD}
     \gamma_{(i}\Dpartial_{j)}\theta^B
              (F^{ik\,C}F_{k}{}^{j\,D} + \quart \eta^{ij} F_{kl}^CF^{kl\,D})
    \nonumber\\
  \label{T2}
  && \quad  +\half\, {\cal A}^{ABCD}
            \gamma_{ijk}
      \{\Dpartial^k\theta^B F^{il\,C} F_l{}^{j\,D}
         - \Dpartial_l\theta^B F^{ij\,C}F^{kl\,D}\}\,.
\end{eqnarray}
The result (\ref{T2}) agrees with the
abelian result (\ref{T2ab}) if we truncate from $U(n)$ to $U(1)$.

The Born-Infeld term now has to reproduce ${\cal T}_2$, while
remaining contributions may be cancelled by
introducing an additional variation of $V_i^A$. It is at this stage that
the choice of $X^\mu$ as a Yang-Mills singlet runs into trouble.
Contributions to this variation of the Born-Infeld term
come from the $F^2$ term in the
action, when the metric $\eta_{ij}$, depending on $X^\mu$, is varied.
Such variations contain a double sum over $U(n)$ indices, i.e.,
they would be of the form (using (\ref{dXnab}))
\begin{equation}
   \partial_iX^\mu \partial_j(\eta^A\Gamma_\mu\theta^A)
       F^{C\,ik} F^{C}{}_k{}^j\,.
\end{equation}
Such terms would have the wrong $U(1)^n$ limit, and cannot be cancelled
by other contributions. Partial integration does not help,
since it produces symmetric second derivatives on $X^\mu$, which
do not occur elsewhere. It is at this stage that we should say
farewell to the embedding coordinates $X^\mu$, and proceed in the static
gauge.

Terms in the Born-Infeld action that might play a role in this analysis are:
\begin{eqnarray}
  {\cal L}_{{\rm BI\,2}} &=& -
     \bigg( \noverm{1}{4}F^{ij\,C}F_{ij}^C
           + \alpha F^{ik\, A}F_k{}^{j\,B}F_{ji}{}^C f^{ABC}
     \nonumber\\
  \label{LBInab2}
     &&\quad
       - \noverm{1}{2} \bar\theta^A {\cal S}^{ABCD}
        \gamma_{(i}\Dpartial_{j)}\theta^B
           \{F^{ik\,C}F_{k}{}^{j\,D} + \quart \eta^{ij} F_{kl}^CF^{kl\,D}\}
     \nonumber\\
     &&\quad
      + \quart\,\bar\theta^A  {\cal A}^{ABCD}
            \gamma_{ijk}
      \{\Dpartial^k\theta^B F^{il\,C} F_l{}^{j\,D}
         - \Dpartial_l\theta^B F^{ij\,C}F^{kl\,D}\} \bigg)\,.
\end{eqnarray}
Note that an $F^3$ term is in principle not excluded in the non-abelian case.

In the static gauge we know how to deal with these terms. Then
we need not vary the $F^2$ term. In the $F^3$ term we have to vary $F$, which
gives an $F^2$ variation with a single $\gamma$-matrix. Therefore it does not
relate to the ${\cal A}$-terms, which have three
$\gamma$-matrices, and we must choose $\alpha$ equal
to zero. In the ${\cal S}$- and ${\cal A}$-terms we can perform partial
integrations to get rid of the $\epsilon$ and $\partial\eta$ terms in
the variation. These give equations of motion of $V$, and can be cancelled
by new variations of $V$. The required identities are:
\begin{eqnarray}
     && \Dpartial_j  (F^{ik\,(C}F_{k}{}^{j\,D)} +
       \quart \eta^{ij} F_{kl}^{(C} F^{kl\,D)})
      = F^{ik\,(C} \Dpartial_j F_k{}^{j\,D)}           \,,
      \\
     &&    \Dpartial_{[k}( F_{i}{}^{l\,[C} F_{lj]}{}^{D]}) -
     \Dpartial_l (F_{[ij}{}^{[C} F_{k]}{}^{l\,D]})
        = - F_{[ij}{}^{[C} \Dpartial_l F_{k]}{}^{l\,D]}    \,.
\end{eqnarray}
The remaining $\eta$ terms in the variation give us precisely ${\cal T}_2$.
This leads to the following new variations of $V_i^A$:
\begin{equation}
\label{delta2Vnab}
  \delta V_i^A =
      +\half\,(\bar\epsilon^B + \bar\eta^B)\,
              {\cal S}^{BCDA} \gamma_k\theta^C F^{ki\,D}
        + \quart\,(\bar\epsilon^B+\bar\eta^B)\,{\cal A}^{BCDA}\gamma_{ikl}\theta^C
                F^{kl\,D}\,.
\end{equation}
Note that the variation of $V_i^A$ no longer agrees with the
result given in (\ref{dVab1}).
However, now we should
compare with the abelian result in static gauge. This gauge choice
requires a compensating worldvolume coordinate transformation, which, when
acting on $V_i$, produces the abelian limit of the ${\cal S}$-contribution
in (\ref{delta2Vnab}). The ${\cal A}$-term in (\ref{delta2Vnab}) vanishes
in the abelian limit.

\section{Summary\label{Summ}}

In this Section we will summarize the results obtained in the
non-abelian case.
The action is the sum of the Born-Infeld and Wess-Zumino terms.
The Wess-Zumino term looks as follows:
\begin{eqnarray}
    {\cal L}_{{\rm WZ}} &=& \epsilon^{i_1\ldots i_{10}} \,
     \bigg\{ \, {1\over 2\cdot 9!}\,\bar\theta^A\sigma_1
   \gamma_{i_1\ldots i_9}\Dpartial_{i_{10}}\theta^A
   \nonumber\\
   &&\quad - {1\over 4\cdot 7!}\,\bar\theta^A \, {\cal P}^{ABC}_{(1)}
      \gamma_{i_1\ldots i_{7}}\Dpartial_{i_{8}}\theta^B
       F^C_{i_9 i_{10}}
   \nonumber\\
\label{LWZall}
   &&\quad\quad + {1\over 16\cdot 5!}\,
   \bar\theta^A \, \big( - \sigma_1 {\cal S}^{ABCD} \big)
      \gamma_{i_1\ldots i_{5}}\Dpartial_{i_{6}}\theta^B
       (F^C F^D)_{i_{7}\ldots i_{10}} \bigg\}\,.
\end{eqnarray}
The Born-Infeld action is:
\begin{eqnarray}
    {\cal L}_{{\rm BI}} &=& -\,\bigg\{
    1 + \half\,\bar\theta^A\gamma^i\Dpartial_i\theta^A
       - \half\, \bar\theta^A \sigma_1 {\cal P}^{ABC}_{(1)}
       \gamma_{[i}\Dpartial_{j]}\theta^B\, F^{ij\,C}
     \nonumber\\
     &&\quad  + \noverm{1}{4} F^{ij\,A} F_{ij}^A
       - \noverm{1}{2} \bar\theta^A
       {\cal S}_{ABCD} \gamma_{(i}\Dpartial_{j)}\theta^B
           \{F^{ik\,C}F_{k}{}^{j\,D} + \quart \eta^{ij} F_{kl}^CF^{kl\,D}\}
     \nonumber\\
\label{LBIall}
     &&\quad    + \quart\bar\theta^A  {\cal A}^{ABCD}
                      \gamma_{ijk}
      \{\Dpartial^k\theta^B F^{il\,C} F_l{}^{j\,D}
         - \Dpartial_l\theta^B F^{ij\,C}F^{kl\,D}\} \,\bigg\} \,.
\end{eqnarray}
In the action we use the following Yang-Mills structures:
\begin{eqnarray}
\label{Pall}
    {\cal P}^{ABC}_{(1)} &=& (i\sigma_2)d^{ABC} \,,
    \\
\label{Sall}
    {\cal S}^{ABCD} &=& {\cal P}^{AE(C}_{(1)}{\cal P}^{BD)E}_{(1)}
                     =  -  d^{AE(C}d^{BD)E}\,,
    \\
\label{Aall}
    {\cal A}^{ABCD} &=&  {\cal P}^{AE[C}_{(1)}{\cal P}^{BD]E}_{(1)}
                      =  -d^{AE[C}d^{BD]E}\,.
\end{eqnarray}
The action is invariant under global supersymmetry transformations
and local $\kappa$-transformations:
\begin{eqnarray}
\label{dthall}
   \delta\bar\theta^A &=& - \bar\epsilon^A+\bar\eta^A\,,
   \\
\label{dVall}
   \delta V_i^A &=& \half\,
      (\bar\epsilon^B+\bar\eta^B)\,\sigma_1{\cal P}^{BCA}_{(1)}\gamma_i\theta^C
      +\half\,(\bar\epsilon^B+\bar\eta^B)\,
            {\cal S}^{BCDA} \gamma_k\theta^C F^{ki\,D}
      \nonumber\\
    &&\quad
     +\quart\,(\bar\epsilon^B+\bar\eta^B)\,{\cal A}^{BCDA}\gamma_{ikl}\theta^C
                F^{kl\,D}\,,
\end{eqnarray}
where the parameters $\epsilon^A$ satisfy the condition
\begin{equation}
   f_{ABC}\,\epsilon^C = 0\,.
\end{equation}
As explained in the previous sections,
the variation of the action  under $\kappa$-symmetry  can be expressed in
terms of
\begin{eqnarray}
  \Gamma^{AB} &=& \Gamma^{(0)}\,\bigg\{
     \sigma_1\delta^{AB}  + {\cal P}^{ABC}_{(1)} \half\gamma^{kl} {F}_{kl}^C
   \nonumber\\
  &&
   - \sigma_1 \,
    {\cal S}^{ABCD} \big( \noverm{1}{8} \gamma_{ijkl}
          F^{ij\,C} F^{kl\,D} -\quart F_{kl}^C F^{kl\,D}  \big)
       \nonumber\\
\label{Gamall}
    &&\quad  -\sigma_1\, {\cal A}^{ABCD}
           \half \gamma_{ij} F^{ik\,C} F_k{}^{j\,D} \bigg\}\,,
   \\
  {\cal T}^A &=& \gamma^i\Dpartial_i\theta^A
         - \sigma_1{\cal P}^{ABC}_{(1)} \gamma_i\Dpartial_j\theta^B F^{ij\,C}
   \nonumber\\
  &&  -{\cal S}^{ABCD}
     \gamma_{(i}\Dpartial_{j)}\theta^B
              (F^{ik\,C}F_{k}{}^{j\,D} + \quart \eta^{ij} F_{kl}^CF^{kl\,D})
    \nonumber\\
\label{Tall}
  && \quad  +\half\, {\cal A}^{ABCD}
            \gamma_{ijk}
      \{\Dpartial^k\theta^B F^{il\,C} F_l{}^{j\,D}
         - \Dpartial_l\theta^B F^{ij\,C}F^{kl\,D}\}\,.
\end{eqnarray}

\section{Gauge fixing\label{Fix}}

In the $\kappa$-symmetric system which we obtained in this paper,
the ordinary supersymmetry is hidden in the local $\kappa$-symmetry, and to
make it explicit, $\kappa$-symmetry should be gauge fixed. This analysis
is very similar to the one done in the abelian case in \cite{Aga}.

To do this analysis it is convenient to write out the $N=2$ doublets
explicitly. We write
\begin{equation}
\Gamma = \left(
  \begin{array}{cc}
     0 & \gamma \\
     \tilde\gamma & 0
  \end{array}   \right)\,,
\end{equation}
with
\begin{equation}
  \Gamma^2 = \left(
    \begin{array}{cc}
       \gamma\tilde\gamma & 0 \\
        0 &  \tilde\gamma\gamma
    \end{array} \right) = \unitmatrixDT \,.
\end{equation}
Here $\gamma$, $\tilde\gamma$ are $32\times 32$ matrices, with
in addition indices $AB$, where $A,B$  run from $1$ to $n^2$.
Then, splitting also the fermions into separate
$N=1$ fermions, we write variations as follows:
\begin{eqnarray}
  \delta\bar\theta_1^A &=& -\bar\epsilon_1^A + \bar\eta^A_1
       \,,\qquad
 \delta\bar\theta^A_2 = -\bar\epsilon_2^A + \bar\eta^A_2
            \,.
\end{eqnarray}
The parameters $\eta$ can be expressed in terms of parameters $\kappa$:
\begin{eqnarray}
  \bar\eta &=& \left(\bar\eta_1\quad \bar\eta_2 \right) =
       \left( \bar\kappa_1+\bar\kappa_2\tilde\gamma
                 \quad
                     \bar\kappa_2+\bar\kappa_1\gamma
                     \right)\,.
\end{eqnarray}
Now we choose a $\kappa$-gauge by setting $\bar\theta_2=0$\,,
which implies that the transformation parameters must satisfy
\begin{equation}
  \bar\kappa_2 = \bar\epsilon_2 - \bar\kappa_1\gamma \,.
\end{equation}
So after $\kappa$-gauge fixing the remaining
$\eta$ is
\begin{equation}
    \bar\eta_1 = \bar\epsilon_2\,\tilde\gamma\,,
\end{equation}
and  the remaining fermions $\chi^A\equiv\theta_1^A$ transform as:
\begin{eqnarray}
   \delta\bar\chi^A &=&
     -\bar\epsilon_1^A +
        \bar\epsilon_2^B \,\tilde\gamma^{BA}
        \,.
\end{eqnarray}
Let us now look at the gauge fixed action. The Wess-Zumino term vanishes
after gauge fixing, since it was off-diagonal in the fermions $\theta_1$
and $\theta_2$. The Born-Infeld term gives:
\begin{eqnarray}
    {\cal L}_{{\rm BI}} &=& - \bigg\{
    1 + \half\,\bar\chi^A\gamma^i\Dpartial_i\chi^A
       + \half\, {d}_{ABC}  \bar\chi^A
       \gamma_{[i}\Dpartial_{j]}\chi^B\, F^{ij\,C}
       + \noverm{1}{4} F^{ij\,A} F_{ij}^A
\nonumber\\
\label{LBIgf}
     &&\quad
       +\ \noverm{1}{2}
       d^{AEC}d^{BDE}\, \bar\chi^A \gamma_{(i}\Dpartial_{j)}\chi^B
           \{F^{ik\,C}F_{k}{}^{j\,D} + \quart \eta^{ij} F_{kl}^CF^{kl\,D}\}
     \\
     &&\quad    -\ \quart  d^{AE[C}d^{BD]E}\, \bar\chi^A
                      \gamma_{ijk}
      \{\Dpartial^k\chi^B F^{il\,C} F_l{}^{j\,D}
         - \Dpartial_l\chi^B F^{ij\,C}F^{kl\,D}\} \,\bigg\} \,. \nonumber
\end{eqnarray}
Note that the
terms of the form $\bar\chi\partial\chi F^2$ are
not symmetric traces of $U(n)$ generators. The symmetric trace is of the form
\begin{equation}
  \tr T_{(A}T_BT_CT_{D)} =
    \noverm{1}{3}\,( d_{ABE}d_{CDE} + d_{CAE}d_{BDE} + d_{BCE}d_{ADE})\,.
\end{equation}
The second line in (\ref{LBIgf}) contains only two of the three contributions
needed for the symmetric trace, the last line contains explicit
anti-symmetrizations
and can be rewritten in terms of structure constants:
\begin{equation}
\label{id3}
   d_{AEC}d_{BDE}-d_{AED}d_{BCE} = f_{ABE}f_{CDE}\,.
\end{equation}
Finally, the linear and nonlinear supersymmetry transformations which leave
(\ref{LBIgf}) invariant are of the form
\begin{eqnarray}
   \delta\chi^A &=& -\bar\epsilon_1^A - \bar\epsilon_2^A
    + \bar\epsilon_2^B\,
    \{  d^{BAC}\,\half \gamma^{kl} F_{kl}^C
    \nonumber\\
     &&\quad    +\ {\cal S}^{BACD}\,
               ( \noverm{1}{8} \gamma_{ijkl}
          F^{ij\,C} F^{kl\,D} -\quart F_{kl}^C F^{kl\,D})
          \,\}   \,,
   \nonumber\\
   \delta V_i^A &=&  -\half\, (\bar\epsilon_1^B - \bar\epsilon_2^B)
            d^{BCA}\gamma_i\chi^C
         -\quart \bar\epsilon_2^B
         d^{BED}d^{ECA} \gamma_{kl}\gamma_i\chi^C F^{kl\,D}
                 \nonumber\\
      &&\quad +\ \half (\bar\epsilon_1^B-\bar\epsilon^B_2)
            S^{BCDA} \gamma_k \chi^C F^{ki\,D}
        \,.
\end{eqnarray}
Note that contributions with ${\cal A}$, the anti-symmetrized product of two
$d$-tensors, do not appear because of (\ref{id3}) and the fact that $\epsilon$
must be in the $U(1)$ direction.

\section{Conclusions\label{Concl}}

In this paper we have obtained the non-abelian generalization of the
Born-Infeld action
up to terms
quartic in the Yang-Mills field strength, and including all fermion
bilinear terms up to terms cubic in the field strength.
The terms of the form $\bar\chi\partial\chi F^2$
violate the symmetric trace conjecture.

The $\kappa$-symmetric construction of the Born-Infeld action involves
the matrix $\Gamma$, satisfying $\Gamma^2=\unitmatrixDT$, which is used to
project away one of the components of the fermion doublet. In the
abelian case $\Gamma$ factorises in a part that is polynomial in $F$, and
the inverse of the Born-Infeld action, which expands to an infinite
series in $F$. In the non-abelian case we are not yet at a stage that
such a factorisation could be recognized. We see however, that the result
(\ref{Gamall}) is consistent with a factorisation of the form:
\begin{eqnarray}
  \Gamma^{AB} &=&
  \Gamma^{(0)}\,\big\{
     \sigma_1\delta^{AC}  + {\cal P}^{ACF}_{(1)} \half\gamma^{kl} {F}_{kl}^F
      - \sigma_1 \,
    {\cal S}^{ACFG} \big( \noverm{1}{8} \gamma_{ijkl}
          F^{ij\,F} F^{kl\,G}  \big)
       \nonumber\\
    &&\quad
      -\ \sigma_1\, {\cal A}^{ACFG}
           \half \gamma_{ij} F^{ik\,F} F_k{}^{j\,G} \big\}
            (\delta^{CB} + {\cal S}^{CBDE}\quart F_{mn}^DF^{mn\,E})\,.
\end{eqnarray}
Note that, as in the abelian case,  ${\cal T}$ (\ref{Tall})
contains the inverse of
the factor that we find in $\Gamma$. Clearly  the second factor in the above
expression is not a $U(n)$ singlet, and therefore
does not correspond to the inverse of the action. Further analysis, which
we plan to do at the cubic and quartic level in $F$, should elucidate in
which sense these factors are related to the Born-Infeld action.

It is intriguing that $\kappa$-symmetry
and worldvolume reparametrisation invariance
appear to be incompatible. Although for
applications such as the construction of
non-abelian BPS states this is not a drawback,
issues of superspace and curved background remain unclear in the static
gauge. One way to try to resolve  the issue of embedding
coordinates would be to look in more detail at the transformation rules
of $V_i^A$. If a formulation with worldvolume reparametrisation
invariance exists, then our formulation should be its gauge fixed version,
and we should recognize the corresponding compensating transformation
in the transformation rule of $V_i^A$. To give an example, let us consider
the possibility that the embedding coordinates are in the adjoint of
$U(n)$, and transform as:
\begin{equation}
   \delta X^{\mu\,A} = d^{ABC}\xi^{i\,B}\partial_i X^{\mu\,C}
            +\delta_\epsilon X^{\mu\,A} + \delta_\eta X^{\mu\,A}\,.
\end{equation}
Let us now gauge fix this extended worldvolume symmetry, by setting
$X^A=0$ for all $A\in SU(n)$, and $X^{\mu\,1}=\delta^\mu_i\sigma_i$
for $A=1\in U(1)$. Then the compensating transformation which
preserves this gauge is of the form:
\begin{equation}
  \xi^A = -\delta_\epsilon X^{\mu\,A} - \delta_\eta X^{\mu\,A} \,.
\end{equation}
Here we have used that $d^{AB1}\sim \delta^{AB}$, in a basis where the
$U(1)$ component of $U(n)$ is labelled by $A=1$.
Then, if the variation of $V_i^A$ under worldvolume coordinate
transformations is of the form
\begin{equation}
    \delta V_i^A = d^{ABC}\xi^{k\, B} F_{ki}^C\,,
\end{equation}
we see that the term
proportional to ${\cal S}$ in (\ref{dVall}) has almost the right form to be
interpreted as a compensating transformation. However, the Yang-Mills
structure in this expression is not quite correct, as the indices of $V$
and $F$ are not on the same $d$-tensor. We have found that if one chooses
in (\ref{P1def}) $c_1=1$, that then the structure comes out all right, and
gives
\begin{equation}
   \delta X^{\mu\,A} \sim d^{ABC}d^{BDE} (\bar\epsilon^D + \bar\eta^D)
        \gamma^k\theta^E\partial_k X^{\mu C}\,.
\end{equation}
It would be interesting to see whether or not the construction of
the generalized Born-Infeld action including the worldvolume
structure indicated above is possible.

As mentioned above, the generalisation to a curved background would be
greatly facilitated by a better understanding of the superspace
structure of the D-brane action. However, there is also another
open issue to consider. Consider
the expression (\ref{calF}), where we give the non-abelian
generalisation of the relation ${\cal F}=F+B$ in a flat
background. In going to a curved background we have to decide
how and where to introduce the NS-NS fields. Should there
be a non-abelian generalisation of the NS-NS field $B$, or
do only the $U(1)$ fields on the worldvolume
couple effectively to the background
fields?\footnote{We thank A.~Tseytlin for a discussion on this issue.}
 Similar questions can be raised about the RR-fields
(see (\ref{LWZnab})),
whose form in a flat background also suggests that a non-abelian
generalisation should be required.

In a future publication \cite{BRS2}, we hope to extend this work to
higher order in $F$, and to apply the results to the
construction of non-abelian BPS states. The simplest
situation to think of is the relation
between  D-branes at angles \cite{BDL} and  overlapping
branes through T-duality \cite{HT}. As was shown in
\cite{DST} the BPS conditions between angles translate to
conditions between magnetic fields $F$ which include
contributions cubic in $F$. Therefore we will have to go at least to
order $F^3$ in the supersymmetry transformation rules to
be able to compare our results with the predictions implied by
\cite{BDL}. In the abelian case the relation between
$\kappa$-symmetric formulations and BPS states was
formulated in \cite{BKOP}. In particular, there it was shown that
the knowledge of $\Gamma$ is in fact sufficient to obtain
BPS states. It would be interesting to generalize these results
to the non-abelian situation.

Finally, it would be instructive to apply other approaches than
the one employed in this paper to find the complete answer. For
instance, one could use the superembedding
techniques developed in \cite{Bandos:1995zw,Howe:1997wf}.
Yet another approach could be
to extend to the non-abelian case the analysis of \cite{Bellucci:2000bd}
where it was shown how the superworldvolume dynamics of 
superbranes can be obtained from nonlinear realizations.

\section*{Acknowledgements}
\noindent
We like to thank
I.~Bandos,
M.~Cederwall,
S.~Ferrara,
S.F.~Hassan,
E.~Ivanov,
R.~Kallosh,
U.~Lindstr\"om,
C.~Nappi,
A.~Peet,
V.~Periwal,
D.~Sorokin,
J.~Troost and
A.~Tseytlin
for useful discussions. Many of these took place at Strings 2000, and
we would like to thank the organizers for providing such a stimulating
environment.
We are grateful to the Spinoza Institute, Utrecht, halfway
between Brussels and Groningen, for the hospitality extended to us.
This work is supported by the European Commission
RTN programme HPRN-CT-2000-00131, in which E.B. and M.d.R. are associated
to the university of Utrecht and A.S. is associated to the university of
Leuven.

\appendix

\section{Properties of $U(n)$ generators, etc\label{AppUn}}

In these notes indices $A,B,\ldots$ run from $1,\ldots n^2$.
We freely raise and lower these indices.

We use the following conventions for Yang-Mills transformations of
the nonabelian Yang-Mills multiplet:
\begin{eqnarray}
   \delta\theta^A &=& f^A{}_{BC} \Lambda^B\theta^C\,,
   \\
   \delta V_i^A &=& -\Dpartial_i\Lambda^A\,,
   \\
   \Dpartial_i\theta^A &=& \partial_i\theta^A + f^A{}_{BC}V_i^B\theta^C\,,
   \\
   F_{ij}^A &=& \partial_iV_j^A-\partial_jV_i^A
     +f^A{}_{BC}V_i^BV_j^C\,,
    \\
   \Dpartial_{[i}\Dpartial_{j]}\theta^A
     &=& \half f^A{}_{BC} F_{ij}^B\theta^C\,.
\end{eqnarray}

The $U(n)$ generators are Hermitian $n\times n$ matrices.
Our normalisation for the trace of two $U(n)$-generators is
\begin{equation}
  \tr T_AT_B = \delta_{AB}\,.
\end{equation}
In general, we write for the product of two $U(n)$ generators:
\begin{equation}
  T_AT_B =  + (d_{ABC} + if_{ABC})T_C\,.
\end{equation}
where $d$ and $f$ are symmetric and antisymmetric in $AB$, respectively.
We recognize that
\begin{eqnarray}
  [T_A,T_B] &=& 2if_{ABC}T_C\,,\nonumber\\
  \{T_A,T_B\} &=& 2d_{ABC}T_C\,.
\end{eqnarray}
{From} this we conclude that
\begin{eqnarray}
  \tr [T_A,T_B]T_C = 2if_{ABC}\,,\nonumber\\
  \tr \{T_A,T_B\}T_C = 2d_{ABC}\,.
\end{eqnarray}
This tells us that in fact $f$ is completely anti-symmetric, and $d$
completely symmetric in $ABC$.

Then we have the Jacobi identity and its generalisations.
These follow from:
\begin{eqnarray}
  &&[[T_A,T_B],T_C] + [[T_B,T_C],T_A]  +[[T_C,T_A],T_B]=0\,,
  \nonumber\\
  &&[\{T_A,T_B\},T_C] + [\{T_B,T_C\},T_A]  + [\{T_C,T_A\},T_B]=0\,,
  \nonumber\\
  &&[T_C,[T_A,T_B]] = \{T_B,\{T_C,T_A\}\} - \{T_A,\{T_B,T_C\}\}\,.
\end{eqnarray}
{From} these we derive the follwoing identities
for the $f$ and $d$ tensors:
\begin{eqnarray}
\label{ff}
  &&f_{ABE}f_{ECD} + f_{BCE}f_{EAD} + f_{CAE}f_{EBD} = 0\,,
  \\
\label{fd}
  &&d_{ABE}f_{ECD} + d_{BCE}f_{EAD} + d_{CAE}f_{EBD} = 0\,,
  \\
\label{ffdd}
  && f_{ABE}f_{ECD} =  d_{CAE}d_{BED} - d_{CBE}d_{AED}\,.
\end{eqnarray}

\end{document}